# Estimating Disaster Resilience of Hurricane Helene on Florida Counties

Reetwika Basu[1,2], Siddharth Chaudhary[1], Chinmay Deval[1,2], Alqamah Sayeed[1,2], Kelsey Herndon[1,2], Robert Griffin[1]

**Abstract**

This paper presents a rapid approach to assessing disaster resilience in Florida, particularly regarding Hurricane Helene (2024). This category four storm made landfall on Florida's Gulf Coast in September 2024. Using the Disaster Resilience Index (DRI) developed in this paper, the preparedness and adaptive capacities of communities across counties in Florida are evaluated, identifying the most resilient areas based on three key variables: population size, average per-person income, and the Social Vulnerability Index (SVI). While the Social Vulnerability Index (SVI) accounts for factors like socioeconomic status, household composition, minority status, and housing conditions—key elements in determining a community's resilience to disasters—incorporating a county's population and per person income provides additional insight. A county's total population is directly linked to the number of individuals impacted by a disaster, while personal income reflects a household's capacity to recover. Spatial analysis was performed on the index to compare the vulnerability and resilience levels across thirty-four counties vulnerable to Hurricane Helene's projected path. The results highlight that counties with high income and lower population densities, such as Monroe and Collier, exhibit greater resilience. In contrast, areas with larger populations and higher social vulnerabilities are at greater risk of damage. This study contributes to disaster management planning by providing a rapid yet comprehensive and reassuring socioeconomic impact assessment, offering actionable insights for anticipatory measures and resource allocation.

[1] Earth System Science Center, University of Alabama in Huntsville, Huntsville, AL, United States
[2] SERVIR Science Coordination Office, NASA Marshall Space Flight Center, Huntsville, AL, United States

# 1. Introduction

Hurricane Helene (2024) struck Florida's Gulf Coast as a powerful Category four storm on September 26, 2024. Following rapid intensification in the Gulf of Mexico, and had strengthened more before its landfall in Florida (Payne and Hollingsworth 2024). Continuous high-speed winds of up to 140 mph and heavy rainfall caused significant damage to life and property across Florida, Georgia, and parts of the southeastern U.S. In anticipation of the storm, the authorities issued mandatory evacuation orders across 33 counties in Florida, with voluntary evacuation options in some (Governor Ron DeSantis, 2024). Florida braced for significant disruptions, with schools, airports, and businesses closing before the storm made landfall. Officials warned that the damage could be catastrophic, particularly in regions along the Big Bend coast. Power outages affected over 800,000 homes and businesses on the eve of September 26 and over 3,500,000 by mid-day September 27 ("US Power Outage Map: Live Outage Data," n.d.) Power outages were also observed in the Black Marble Nighttime at sensor radiance (Day/Night Band) [Figure 1].

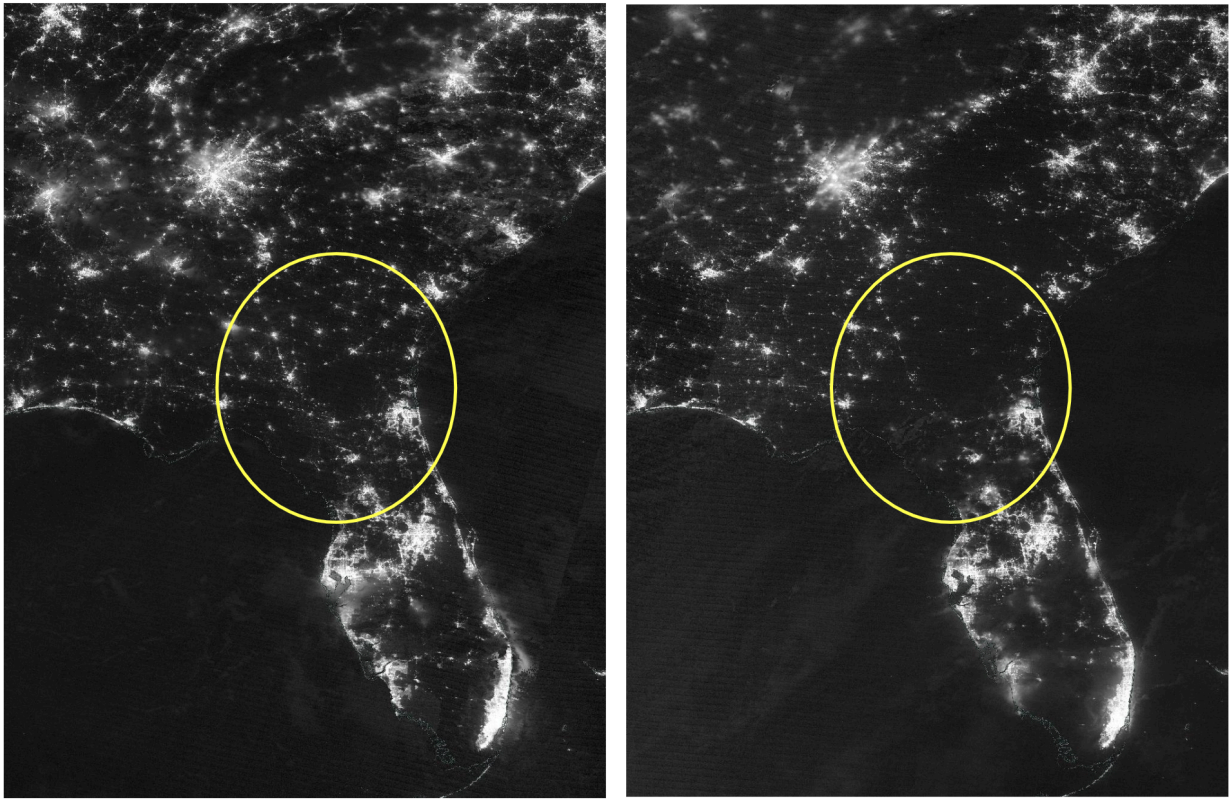

Figure 1: Pre (left) and Post (right) Hurricane Nighttime sensor radiance. [Source: Visible Infrared Imaging Radiometer Suite (VIIRS) aboard the joint NASA/NOAA NOAA-20 (JPSS-1) satellite].

Helene's impact was projected to extend beyond Florida [Figure 2], with warnings of flash floods, damaging winds, and heavy rainfall expanding into Georgia, Alabama, and the South and North Carolinas. Historically, more strong and more destructive hurricanes have impacted Florida more than any other state (Malmstadt, Scheitlin, and Elsner 2009). Since 2000, Florida has endured Hurricanes Michael (category four), Charley, Irma, and Ian (category four), which have caused massive destruction (National Hurricane Center 2019). For example, Hurricane Ian, in 2022, is estimated to have caused damage of $118.5 billion, followed by $50 billion in damage caused by Hurricane Irma in 2017 (NOAA, n.d.). The Florida counties most affected over the year, identified by FEMA (FEMA, n.d.a) as being on the direct path of hurricanes, are Monroe, Bay, Lee, Charlotte, and Collier. Hurricane Helene is being compared to Hurricane Michael (2018) and Hurricane Idalia (2023), which were massively destructive in the Gulf Coast region. Timely and informative metrics to inform evacuation and preparedness strategies are crucial, as decision-makers must quickly allocate limited resources in response to the potential impacts of these intense events.

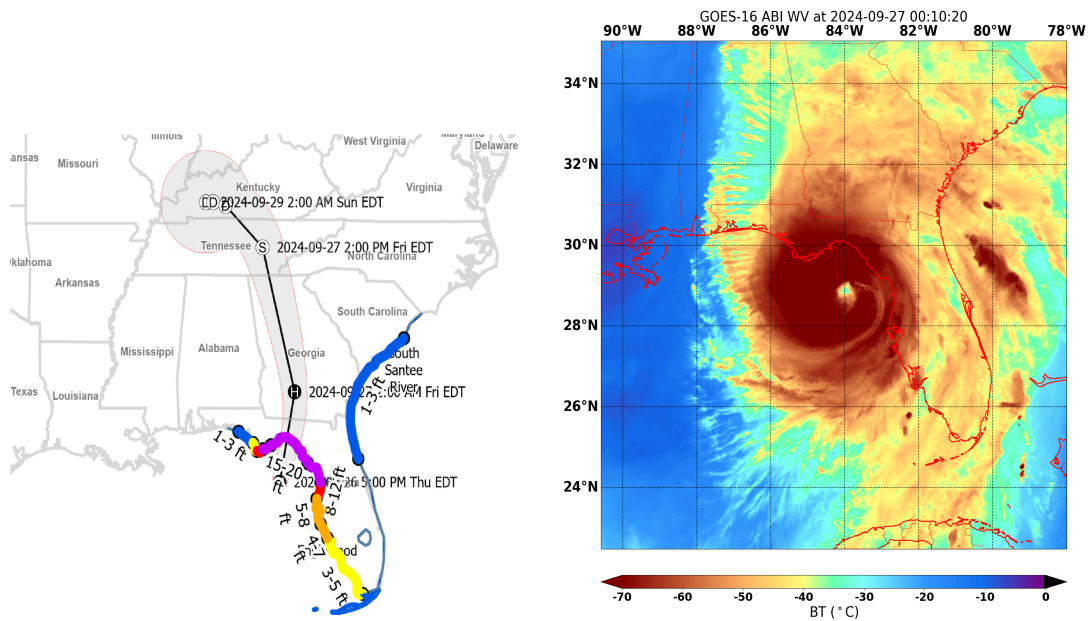

Figure 2: (a) Track and cone of hurricane Helene and storm surge provided by National Hurricane Center NOAA (b) GOES-16's Advanced Baseline Imager (ABI) captures Water Vapor (WV) imagery, which measures radiation emitted by water vapor in the atmosphere. It provides insight into the movement of moisture and air masses in the mid to upper troposphere pre and post-hurricane Helene. This imagery shows weather patterns, storm development, and circulation features.

Several studies have developed conceptual frameworks that contribute to the discourse on disaster resilience (Manyena 2006; Adger et al. 2005). These studies are primarily categorized into those that address resilience based on adaptive capacities (Ross 2013; Zukowski 2014; Bruneau et al. 2003) and those that consider vulnerability (Cutter, Boruff, and Shirley 2003; Manyena 2006) as the primary contributor to measuring resilience. Adaptive capacities refer to the community's ability to cope with disasters in terms of available resources and infrastructure (Zukowski 2014; Adger 2003). Conversely, vulnerability accounts for the community's socioeconomic, environmental, and physical factors (Cutter, Boruff, and Shirley 2003; Blaikie et al. 2004). While adaptive capability contributes positively to disaster resilience, vulnerability has the opposite effect (Rifat and Liu 2020). Some previous studies have focused on demographic details and direct economic losses, overlooking the social vulnerability factor that plays a significant role in determining efficient disaster management methods (Khalid and Ali 2020). The discourse around defining risk and resilience is complex, with researchers presenting various interpretations. Some view resilience as the inverse of risk, while others consider it a strategy to manage risk. Several studies have explored the interplay between these concepts as both distinct yet related (Yu et al., 2023).

A comprehensive approach to describing disaster resilience is crucial, as it allows for a more nuanced understanding of the issue and, therefore, the development of a more effective response. The Disaster Resilience Index (DRI) described in this paper takes a comprehensive approach, considering variables associated with both adaptive capabilities and vulnerability. In the era of burgeoning AI and machine learning models, using straightforward Geographical Information Systems (GIS) methods combined with simple socio-economic models is of great significance in enabling faster estimations of the vulnerability of regions and facilitating their use by non-experts (Kemper and Kemper 2020). In this paper, a rapid socioeconomic impact assessment aims to provide quick, actionable insights for informed anticipatory actions or disaster management strategies to mitigate damage. This study demonstrates a method to deliver fast metrics using widely available demographic data, such as income, population, and infrastructure, rather than conducting more extensive economic analyses that require more time and resources.

The DRI proposed in this study accounts for average personal income, population within each county, and social vulnerability. These are the three most essential indicators observed in separate studies [Cutter 2003; Cutter 2016; Liu et al. 2020] that have historically been used to identify risk or resilience. Population size significantly impacts disaster outcomes and the ability to recover from disaster when combined with personal income. Personal income represents the direct ability of the population to address the shocks and disasters from a particular event. Unlike other income-related indexes like the poverty rates that address the economic capability across a population segment, personal income accounts for individual capacities that help people recover even without appropriate insurance. This early application of the DRI demonstrates how it can be rapidly deployed to aid in the identification of Florida's most vulnerable regions/counties facing catastrophic events like Hurricane Helene. The DRI also estimates the degree of climate resilience in each county on Hurricane Helene's path. The results from this study can help improve disaster management planning and response in the most vulnerable and storm-exposed areas of Florida.

## 2. Methods

### 2.1 Data and Sources

Calculating the DRI requires county-level population, income, and Social Vulnerability Index data. The data used to identify the Social Vulnerability Index (SVI) and Population per county are sourced from the Centers for Disease Control and Prevention (CDC). The most recent data available for this analysis were collected in 2022. The CDC SVI is a robust dataset designed to identify socially vulnerable communities across the United States (CDC-ATSDR 2020). SVI incorporates 16 critical variables from the U.S. Census Bureau's American Community Survey, categorized into four themes (see Figure 3): socioeconomic status, household composition, minority status/language, and housing/transportation (CDC-ATSDR 2024). The index provides vulnerability scores at the census tract and county levels, enabling detailed social vulnerability analysis across varying geographic scales (Tran et al. 2023). Updated periodically, the SVI is a valuable tool for public health officials, researchers, and policymakers to identify areas that may require additional support during emergencies or public health crises ("CDC SVI Documentation 2020 | Place and Health | ATSDR" 2022; "CDC/ATSDR Social Vulnerability Index (CDC/ATSDR SVI)" 2024) and for disaster preparedness. The Agency for Toxic Substances and Disease Registry (ATSDR), Centers for Medicare and Medicaid Services (CMS), and other nonprofits and NGOs like the American Red Cross use this dataset to assess the resilience of communities facing disasters.

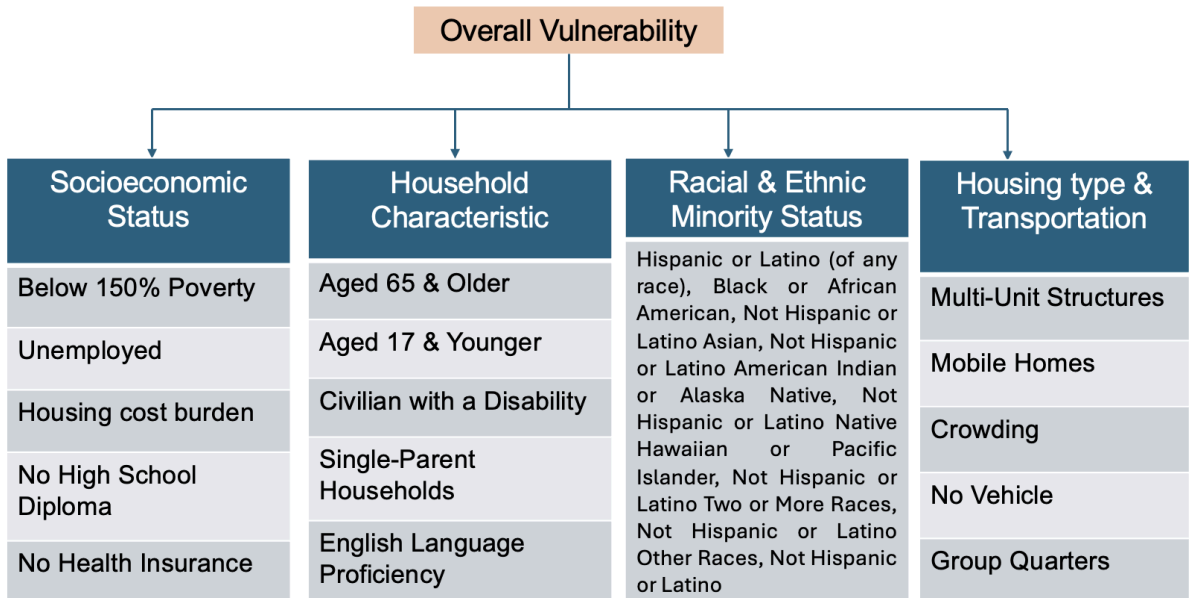

Figure. 3: Social Vulnerability Index (SVI) [Source: CDC SVI 2022 Data Directory]

The county-level average per person income data used in this study is from the Bureau of Economic Analysis (BEA). The BEA is the federal agency responsible for providing information on vital economic indicators, including U.S. gross Domestic Product (GDP), state and local level economic data, foreign trade statistics, and industry-specific reports. Personal income data is one of its core reporting variables, including income from wages, individuals, interests, and government benefits, segregated by region.

## 2.2 Disaster Resilience Index

Disasters do not impact all regions and communities equally. Numerous studies (Morrow 1999; Fothergill and Peek 2004; Flanagan et al. 2011) have shown that factors such as population sizes, personal incomes, and demographic characteristics—such as age, employment status, education level, and housing conditions—play a significant role in determining the level of impact. This paper uses population size, personal income, and SVI variables to define the Disaster Resilience Index (DRI). For this study, we define resilience as an individual or community's capacity to recover from a disaster's impacts. These findings have practical implications for disaster management and community resilience, as they can guide resource allocation and intervention strategies. DRI Index is defined as:

$$DRI_i = 0.33[\frac{P_i - P_{min}}{P_{max} - P_{min}} + (1 - \frac{I_i - I_{min}}{I_{max} - I_{min}}) + SVI_i],$$

$i$ indicates each county, $P$ indicates the population of the specific county ($i$), $P_{min}$ represents the minimum population amongst all Florida counties, whereas $P_{max}$ refers to the maximum population. $I$ refer to average personal income. $I_{min}$ Indicates the lowest average personal income amongst all Florida counties and $I_{max}$ is the maximum average individual income. $SVI$ is the social vulnerability index obtained from CDC. The $SVI$ (ranging from 0 to 1) calculation for each county accounts for the socio-economic status, household characteristics, racial and ethnic minority status, housing type and transportation [Fig.2]. A high SVI closer to 1, signals that the community is more likely to suffer severe impacts alongside counties with high populations susceptible to damage.

Normalization of the population and average personal income per county ensures that each factor contributes equally and proportionally to the damage index. This prevents the assumption that higher population density in a region will automatically lead to more significant damage. A lower average income reflects a lower capacity for recovery from hurricane damage. The normalized per-person income is subtracted from 1 to represent that appropriately. Equal weight has been assigned to each variable in the damage index to create a balanced approach, ensuring that all factors contribute equally to assessing the overall impact. DRI values can range from 0 to 1; a higher DRI indicates that a community is better prepared to withstand and recover from the hurricane, possibly with an adequate response system and solid capacities in place.

## 3. Results and Discussion

The DRI was calculated for each county along the path of Hurricane Helene (NOAA, n.d.) using the methods and data in section 2. Results were visualized in a GIS to show the extent of disaster resilience in each county, with the data categorized into five classes calculated using the equal quantiles method for qualitative analysis. Figure 4 shows the estimated DRI for each county (Figure 4d) along with the corresponding population (Figure 4a), income (Figure 4b), and SVI (Figure 4c). The counties identified by NOAA as being in the storm's path were Charlotte, Citrus, Collier, Dixie, Franklin, Gadsden, Gulf, Hernando, Highlands, Hillsborough, Jefferson, Lee, Leon, Levy, Liberty, Madison, Manatee, Monroe, Pasco, Pinellas, Sarasota, Taylor, and Wakulla. Collier, Monroe, Sarasota, and Charlotte counties have a very low DRI, indicating they are the least resilient to hurricane damage. These are followed by Gulf, Citrus and Hernando had relatively low DRI, and Lee, Pinellas, Pasco, Levy, and Leon had a moderate ability to deal with the disaster. Figure 4c shows Sarasota and Charlotte have very low SVI, and Collier and Monroe have moderate SVI scores. The population in the counties mentioned above is very high or relatively higher than other counties on the storm path (Figure 4a), with relatively high personal incomes compared to the rest of the counties (Figure 4b). Franklin, Liberty, Gadsden, Hillsborough, and Highlands, on the other hand, have the highest DRI of all counties on the storm path.

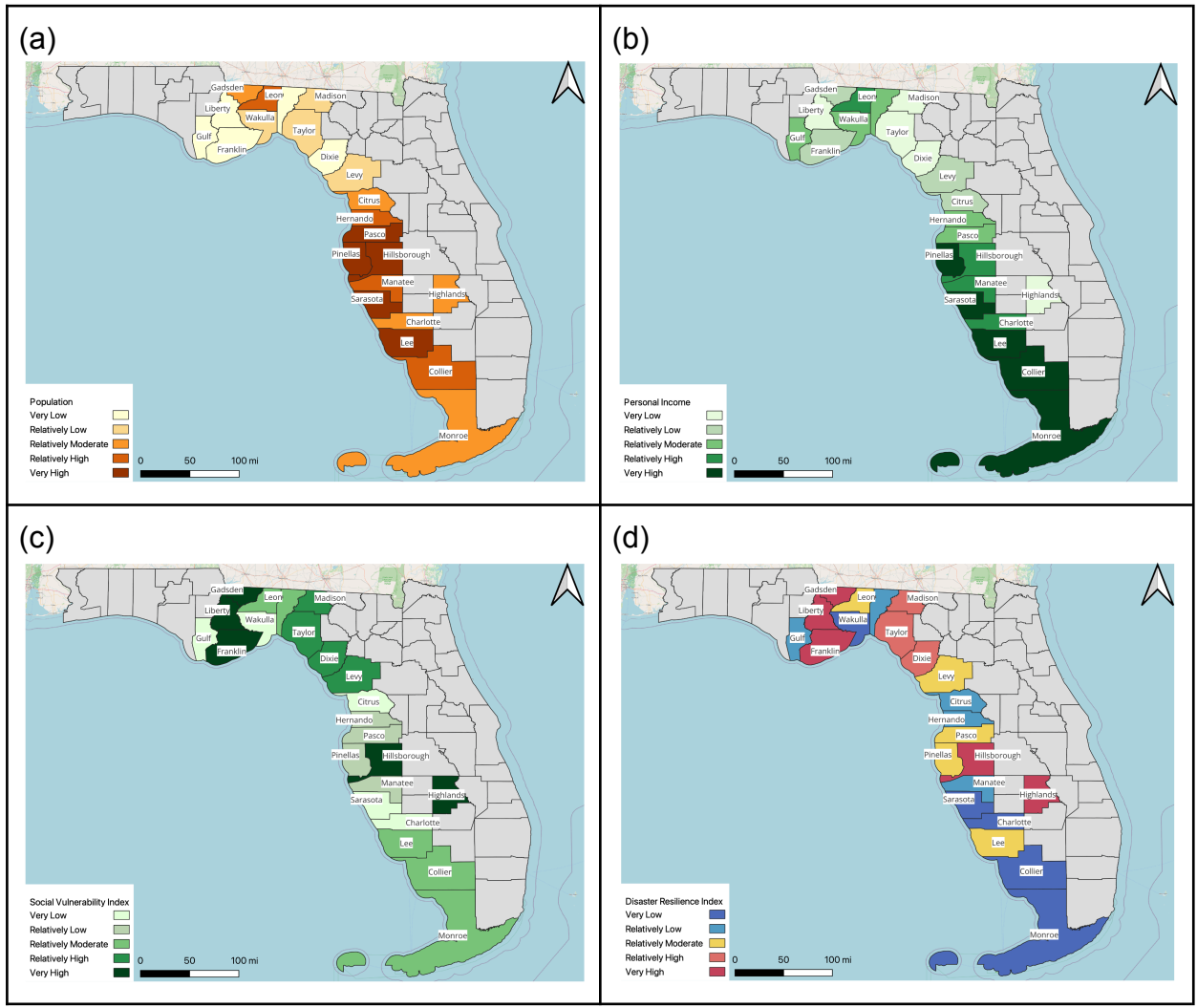

Figure 4: Comparison of Demographic Factors and Adaptive Capacities (a) Population (b) Personal Income (c) Social Vulnerability Index of counties on Hurricane Helene's path and (d) Calculated Disaster Resilience Index

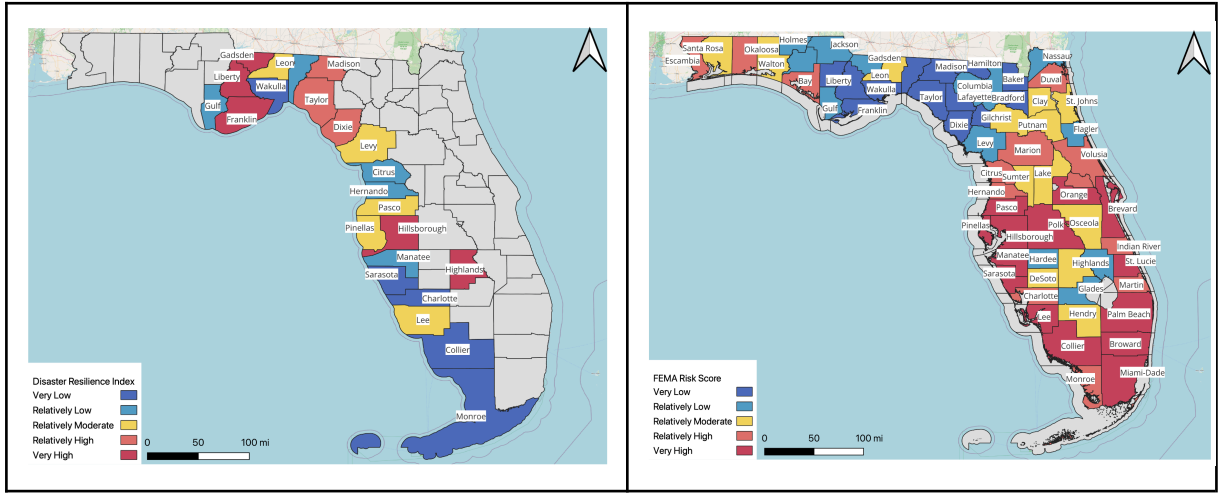

Figure 5: Comparison of DRI (left) and FEMA risk index (right)

The FEMA National Risk Index (NRI) risk index is a widely used tool by national and local governments for disaster preparedness. FEMA's Risk Index is formulated as a ratio of expected annual loss in the county multiplied by social vulnerability, and community resilience. FEMA's social vulnerability is based on the CDC SVI. However, the community resilience data is sourced from the Baseline Resilience Indicators for Communities (BRIC) Index by the University of South Carolina's Hazards and Vulnerability Research Institute. The expected annual loss is calculated from expected and historical exposure to disasters. A lower NRI typically corresponds to a higher DRI, and vice versa. However, a high NRI, as seen in places like Lee County, doesn't automatically indicate disaster vulnerability. The higher income levels of Lee County's residents enable faster disaster recovery, which is reflected in the county's moderate DRI value. The Disaster Risk Index (DRI) developed in this paper differs from the FEMA NRI in that it assigns equal weight to both social and economic factors, making it more adaptable to specific regions—particularly those where social and economic disparities play a significant role in shaping disaster outcomes. The FEMA NRI provides a broader assessment of natural hazard exposure, factoring in expected losses, which are intensified by social vulnerability and mitigated by community resilience. However, it does not directly account for personal income levels in any of its indicators. In contrast, the DRI incorporates personal income, making it a more straightforward tool for implementation.

The Federal Emergency Management Agency (FEMA, n.d.b) identified Collier, Lee, Sarasota, Manatee, Pasco, Hillsborough, and Pinellas as having a very high-risk Index, followed by Monroe, Charlotte, Hernando, and Citrus (all have very low, relatively low or relatively moderate DRI's except Hillsborough). On the other hand, Wakulla, Dixie, and Liberty have relatively low-risk indexes [Figure 5]. Hilborough has a very high SVI and population but also demonstrates a very high average personal income. The personal income contributes immensely to the county's ability to deal with a disaster but at the same time certain categories in population might be bearing high risks driving the risk index very high.

The key findings of our study reveal a high DRI for Monroe and Collier, followed by Sarasota and Wakulia. The high-risk indexes of Monroe and Charlotte can be attributed to several factors, including high expected annual loss due to their geographic location, exposure to hurricanes, and elevated social vulnerabilities among specific population categories. It's important to note that high income and available resources may not be strong enough to offset the damage caused by the hurricane. Even though Monroe and Collier are relatively affluent with moderate SVI scores, their population density and preparedness to address hurricane impacts may be lacking in certain demographic groups.

Normalizing income in the DRI calculation reduces reliance on income alone, illustrating that wealth does not guarantee recovery from damages. This indicates that Monroe and Collier counties face a 'double-edged sword'. While higher income might imply a capacity for quicker recovery, their relatively moderate social vulnerability and lower population density ultimately diminish their DRI. At the same time, frequent and intense storms contribute to high expected annual losses, further elevating their FEMA Risk Index. These combined factors render them both highly vulnerable and less resilient, underscoring the importance of integrating resilience-focused metrics like the DRI to enhance disaster preparedness and recovery strategies. These findings align with recent research that advocates for a greater emphasis on resilience rather than focusing solely on need or vulnerability (Manyena 2006).

## 4. Conclusions

The Disaster Resilience Index (DRI) serves as a practical tool for assessing a county's capacity to recover from disasters, contrasting with FEMA's National Risk Index, which focuses primarily on quantifying total annual losses. Unlike the FEMA NRI, which emphasizes the expected damages and overall risk, the DRI offers a more focused evaluation of a community's adaptive capacity and ability to "bounce back" after a disaster. This makes the DRI particularly useful for guiding immediate resource allocation and anticipatory actions, ensuring that communities most vulnerable to long-term impacts receive the necessary support.

The DRI's advantage lies in its rapid assessment capability, offering actionable insights that policymakers can use to allocate resources effectively. While the FEMA NRI is valuable for broader risk assessment, the DRI's granular focus on population size, income, and social vulnerability provides a more targeted approach to assessing recovery potential. This allows for a proactive disaster management strategy, focusing not just on risk but also on the ability of communities to recover, thus fostering long-term resilience.

Although the DRI may not encompass all relevant factors—such as storm intensity, evolving demographic metrics, storm surge, and distance from landfall—it provides a simplified yet insightful framework for identifying the most vulnerable communities. Enhancing the DRI by integrating dynamic vulnerability factors such as proximity to storm tracks, wind speeds, and coastal areas would further refine this tool, making it more comprehensive and improving its predictive power for disaster resilience.

While comprehensive in assessing risk and potential loss, FEMA's NRI lacks a focus on post-disaster recovery and resilience capacity, which is where the DRI excels. Together, these tools are complementary in disaster planning. By combining the broader risk assessment of the NRI with the resilience-focused insights of the DRI, policymakers can adopt a more holistic approach that addresses both immediate risks and long-term recovery potential.

Ultimately, our findings underscore the importance of incorporating resilience-focused metrics into disaster management strategies. By shifting the focus from purely quantifying risk to evaluating resilience and recovery capacity, the DRI empowers policymakers to equip communities to better withstand and recover from future disasters. However, the current iteration of the DRI represents a rapid deployment of this new index. Future research and robust statistical analysis are essential to refine its methodology, ensuring the tool gains greater trust and reliability. As the DRI is validated and improved, its utility and accuracy for practical disaster risk reduction efforts will become even more impactful.

# Acknowledgements

The authors are grateful to the Lab of Applied Sciences (LAS) in the Earth System Science Center (ESSC) at the University of Alabama in Huntsville (UAH) for their invaluable support and resources that made the research possible in such a short time. The collaborative environment greatly contributed to the rapid assessment conducted in this study.

# Data Availability Statement

All data used in this study are available online. The calculated DRI can be downloaded from https://zenodo.org/records/13883059.